\documentclass[a4paper]{article}

\usepackage{INTERSPEECH2020}
\usepackage{cite}
\usepackage{comment}
\usepackage{color,soul}

\title{Assessment of Parkinson's Disease Medication State through Automatic Speech Analysis}
\name{Anna Pompili$^1$, Rub\'{e}n Solera-Ure\~{n}a$^1$, Alberto Abad$^{1,2}$, Rita Cardoso$^{3,5,6}$, Isabel Guimar\~{a}es$^{5,6,7}$, Margherita Fabbri$^8$, Isabel P. Martins$^{4,5}$, Joaquim Ferreira$^{3,5,6}$\thanks{This work has been supported by national funds through Funda\c c\~ao para a Ci\^{e}ncia e a Tecnologia (FCT) with reference UIDB/50021/2020.}}
\address{
  $^1$INESC-ID, Lisbon, Portugal \\
  $^2$Instituto Superior T\'{e}cnico, Universidade de Lisboa (UL),  Portugal \\
  $^3$Laboratory of Clinical Pharmacology and Therapeutics, Faculdade de Medicina, UL,  Portugal \\
  $^4$Laborat\'{o}rio de Estudos de Linguagem, Faculdade de Medicina, UL, Portugal \\
  $^5$Instituto de Medicina Molecular, Lisbon, Portugal \\
  $^6$CNS - Campus Neurol\'{o}gico S\'{e}nior, Torres Vedras, Portugal \\
  $^7$Alcoit\~{a}o Schoool of Health Sciences, Santa Casa da Miseric\'{o}rida de Lisboa, Portugal\\
  $^8$ Departments of Clinical Pharmacology and Neurosciences, CIC 1436, NS-Park/FCRIN network and NeuroToul COEN Center, INSERM, CHU of Toulouse and University of Toulouse, France
  }
\email{anna@hlt.inesc-id.pt, rsolera@hlt.inesc-id.pt, alberto.abad@inesc-id.pt}

\begin{document}

\maketitle
\begin{abstract}
  Parkinson's disease (PD) is a progressive degenerative disorder of the central nervous system characterized by motor and non-motor symptoms. As the disease progresses, patients alternate periods in which motor symptoms are mitigated due to medication intake (ON state) and periods with motor complications (OFF state). The time that patients spend in the OFF condition is currently the main parameter employed to assess pharmacological interventions and to evaluate the efficacy of different active principles. In this work, we present a system that combines automatic speech processing and deep learning techniques to classify the medication state of PD patients by leveraging personal speech-based bio-markers. We devise a speaker-dependent approach and investigate the relevance of different acoustic-prosodic feature sets. Results show an accuracy of 90.54\% in a test task with mixed speech and an accuracy of 95.27\% in a semi-spontaneous speech task. Overall, the experimental assessment shows the potentials of this approach towards the development of reliable, remote daily monitoring and scheduling of medication intake of PD patients.
\end{abstract}

\section{Introduction}
\label{sec:intro}

Parkinson's disease (PD) is a progressive degenerative disorder of the central nervous system characterized by motor and non-motor symptoms. The cardinal motor signs of PD include the characteristic clinical picture of resting tremor, rigidity, bradykinesia, and impairment of postural reflexes, while non-motor symptoms include cognitive disorders, and sleep and sensory abnormalities. Motor symptoms of PD influence also the speech production of language. Dysarthria, which is characterized by a weakness, paralysis, or lack of coordination in the motor-speech system, is typically observed in PD patients and affects respiration, phonation, articulation and prosody. As a consequence, the main deficits of PD speech are loss of intensity, monotony of pitch and loudness, reduced stress, inappropriate silences, short rushes of speech, variable rate, imprecise consonant articulation and harsh and breathy voice. Both motor symptoms and speech impairments slowly worsen during the disease with a nonlinear progression. 
%Speech and voice impairments also deteriorate along the curse of the disease, 
Patients in advanced stages typically present more severe speech abnormalities, with voice disorders and articulatory deficits being the most prevalent symptoms. At the final stage of the disease, articulation is frequently the most impaired feature~\cite{ho1999speech,sapir1999speech,Logeman1978}. As the disease progresses, patient alternate periods in which motor symptoms are mitigated due to medication intake (ON state), and periods with motor complications (OFF state). The time that patients spend in the OFF condition is currently the main parameter used to assess pharmacological interventions. %and to evaluate the efficacy of different active principles. 
%Currently, the only available method to collect such information consists of self-reporting diaries

In this scenario, it is of paramount importance to investigate whether speech characteristics can be used as a bio-marker to monitor and track ON-OFF conditions of PD patients. If this would be possible, then speech cues may bring further insights into the dynamics of the disease over time, not only by providing accurate information on medication states and alterations, but also by investigating its correlation with motor fluctuations. 
However, while the effects of dysarthria in PD have been deeply investigated, both in drug-naive state and under pharmacological treatment~\cite{skodda2011aspects,bocklet2011detection,rusz2011quantitative,whitfield2018fluency,skodda2011vowel,hammen1996speech},
it is still not clear whether speech analysis could be helpful to differentiate between ON-OFF states induced by pharmacological intake. In fact, the clinical literature presents controversial results on the effect of Levodopa on speech abilities~\cite{okada2016effects,poluha1998handwriting,skodda2011intonation,de2006levodopa,anderson1999developmental,im2019effect}, while no computational studies addressing this topic have been identified in the literature.

To the best of our knowledge, this is the first work that investigates the problem of automatically identifying medication states of PD patients from speech. We propose a system that combines speech processing and deep learning techniques to perform automatic classification of the medication state by leveraging personal speech-based bio-markers. We devise a speaker-dependent experimental setup that uses the FraLusoPark corpus\cite{fralusoparkDB}. This dataset contains recordings from 74 native Portuguese PD patients performing several vocal tasks, both in drug-naive state and under pharmacological treatment. The results show that it is possible to classify ON-OFF condition with an accuracy of 90.54\% using speaker-dependent models. Moreover, it was found in previous work that the type of vocal task has a relevant impact on the capability of automatic approaches to characterize speech impairments in PD patients~\cite{pompili2017automatic,bocklet2013automatic,rusz2011quantitative}. When we extend such a per-speech task analysis to this work, we find that the accuracy of the system improves to 95.27\% on a semi-spontaneous storytelling task.
These results are of extreme relevance, not only because they represent a first contribution to the automatic assessment of the medication state of PD patients. They also confirm that it is possible to obtain reliable  models using a relatively reduced amount of patients' speech data. Altogether, we consider that this is a relevant achievement towards the development of reliable, personalized and non-invasive health solutions for the daily monitoring and scheduling of medication intake of PD patients. This technology could be deployed at patients' homes and remotely provide the clinicians with constant, updated information.

The rest of this work is structured as follows: Section \ref{section:SOA} introduces the relevant state of the art. In Section \ref{section:corpus} and \ref{section:methods}, we present the dataset used in this study and a description of our methodology. Experimental results are presented in Section \ref{section:exp} and the paper ends with conclusions in Section \ref{section:conclusions}.

\section{Related work}
\label{section:SOA}

The effects of Levodopa treatment on speech abilities have been analyzed considering different dimensions of voice quality. Several studies focused their attention on phonatory and articulatory abilities. Goberman \textit{et al.}~\cite{goberman2002phonatory} found small differences between groups of patients before and after medication, which were pointing to phonatory improvements.
Jacobi \textit{et al.}~\cite{jacobi2019effect} analyzed the Vowel Articulation Index (VAI) in 10 participants of different languages, finding that VAI was not significantly affected by pharmacological treatment. On the other hand, Okada \textit{et al.}~\cite{okada2016effects} reported significantly expanded Vowel Space Area (VSA) after Levodopa treatment of 21 PD patients, contrarily to the findings by Poluha \textit{et al.}~\cite{poluha1998handwriting} on a corpus of 10 subjects. Some authors shifted their attention towards the analysis of prosodic deviations. Using a reading task composed of four complex sentences, Skodda \textit{et al.}~\cite{skodda2011intonation} analyzed intonation and speech rate in a group of 138 patients. The authors found significant changes from the first to the last sentence of the passage, but generally, no effect of Levodopa administration was found on the reading of the entire task. In this respect, De Letter \textit{et al.}~\cite{de2006levodopa} found no differences in the speech rate of 25 patients reading a text passage in OFF and ON states. %Add Ho?
Finally, when investigating the effects of Levodopa on developmental stuttering, Anderson \textit{et al.}~\cite{anderson1999developmental} reported an increase of disfluencies in the spontaneous speech of 1 patient during the ON period. This is in disagreement with the findings of Im \textit{et al.}~\cite{im2019effect}, which examined 51 patients while reading a standard passage. In fact, the authors found an improvement of disfluency in the ON state that was proportional to the severity of speech impairments presented in the OFF state. The discrepancies found in the studies described above may depend on many factors, like different methodological approaches or the small number of samples observed in some studies. Overall, it is rather consensual that Dysarthria is considered to be responsive to treatments. Nevertheless, there exists also the hypothesis that pharmacological intake may induce additional speech deficits caused by side effects of the treatment itself~\cite{Moreau_misconception}.

While speech and language technologies may potentially provide valuable contributions to the analysis of patients' states, to the best of our knowledge, this is the first study that investigates a computational approach to this problem. In this respect, our work brings the following contributions. First, it is a fully-automatic method purely based on acoustic bio-markers of speech. Second, we explore the use of deep architectures in a challenging domain like the health area, characterized by relatively constrained datasets with respect to the ones typically used with these models. Third, by exploiting a relatively large corpus composed of 74 PD patients performing several tasks, we show that is possible to achieve reliable personalised models to identify ON-OFF medication states, particularly in semi-spontaneous vocal production tasks. 

\section{Corpus description}
\label{section:corpus}

The corpus used in this study is the FraLusoPark database \cite{fralusoparkDB}, which contains the recordings of 139 European Portuguese speakers. The control group, composed of 65 healthy volunteers, is age-matched and sex-matched with the PD group, composed of 74 subjects. The patient group contains 38 male with an average age of 65 years ($\pm$11.9), and 37 female, with an average age of 70 ($\pm$8.5) years. Each patient was recorded twice on the same day, OFF medication (i.e.: at least 12 hours after withdrawal of all anti-Parkinsonian drugs), and ON medication (i.e.: following at least 1 hour after the administration of the usual medication). All the recordings where performed under exactly the same environmental conditions. 

Participants were required to perform 9 speech production tasks with an increasing complexity in a fixed order: (1) three repetitions of the sustained phonation of the vowel /a/, (2) two repetitions of the maximum phonation time (vowel /a/ sustained as long as possible), (3) oral diadochokinesia (repetition of the pseudo-word \textit{pataka} at a fast rate for 30 s.), (4-5) reading aloud of 10 words and 10 sentences, (6) reading aloud of a short text ("The North Wind and the Sun"), %\cite{NorthWind}, 
(7) reading aloud of a set of sentences with specific prosodic properties,
(8) storytelling speech guided by visual stimuli, %~\cite{Frog}
and (9) engaging in a spontaneous conversation for around three minutes. 
\begin{comment}
Thus, the dataset contains 9 ON and 9 OFF recordings for every speaker, with a total of 1332 recordings.
\end{comment} 
Table \ref{tbl:corpus_clean} shows the duration of the corpus by task and medication state after removal of pauses and therapists' interventions.

\begin{table}[t!]
\centering
\caption{Duration by task and medication state of the FraLusoPark dataset (hh:mm:ss) after automatic pre-processing.}
\label{tbl:corpus_clean}
\begin{tabular}{lcc}
\hline
\textbf{Task}              & \textbf{ON} & \textbf{OFF} \\ \hline
/a/                        & {00:31:22} & {00:35:22} \\
MPT                        & {00:39:22} & {00:38:24} \\
DDK                        & {00:32:36} & {00:30:50} \\
Reading 10 words           & {00:10:34} & {00:09:52} \\
Reading 10 sentences       & {00:18:24} & {00:17:56} \\
Reading text               & {00:40:04} & {01:27:56} \\
Reading prosodic sentences & {01:22:21} & {00:41:08} \\
Story telling              & {00:37:59} & {00:41:14} \\
Conversation               & {02:40:38} & {02:51:21} \\ \hline
\textbf{Total}            & 07:18:17  & 07:36:19 \\ \hline
\end{tabular}
\end{table}

\section{Methodology}
\label{section:methods}

Our approach to the automatic classification of ON-OFF states relies on three main stages. The first one deals with the automatic removal of non-speech segments and therapists' interventions. Then, acoustic-prosodic markers are computed from the clean speech signals. Finally, PD patients' medication state is modeled with deep neural networks.

\subsection{Pre-processing and segmentation}

We describe here our automatic approach to remove speech segments corresponding to silent pauses and instructions of the speech language therapists. Each audio file in this corpus corresponds to the continuous, unsegmented recording of a task. In our previous work with this dataset~\cite{pompili2017automatic}, the recordings of the control group were processed semi-automatically to generate segmentation annotations with information of speech/non-speech segments and speaker (patient/clinician) turn. These data are exploited to generate the same kind of annotations for the Patient recordings, using a fully-automatic approach based on speech/non-speech segmentation (SNS) followed by speaker identification. 
%The recordings of the patient group in the OFF state were not considered in our previous work and now we have to approach the same issue. However, as a result of our previous work, we are now provided with the information of the speech segments that belong to the therapists in these two groups of the corpus.
The SNS segmentation is based on a simple finite state machine encoding  heuristic rules about the minimal segment duration and state-changing thresholds, and takes as input the likelihoods provided by a bi-Gaussian model of the log-energy distribution computed for each recording. In the speaker identification step, we follow the classical GMM-MAP modeling approach\cite{reynolds2000speaker}, in which a universal background model (UBM) is first trained and then adapted to a target speaker using some enrolment data. In this case, we train the UBM using all the speech segments in the control group recordings (07h 19m 27s) and a therapist model using the therapist speech segments in the control group recordings (07m and 40s). Finally, for each speech segment in the Patient recordings, the  therapist vs. UBM likelihood ratio is computed and compared to a decision threshold $\Theta$. In this work, we are specially concerned with the removal of the therapist's speech segments, while, up to some extent, we can afford to eliminate some  data of patient's speech. For that reason, the decision threshold was empirically set to minimize the therapist insertion errors according to experiments conducted  in the Control group recordings.

\begin{comment}
The threshold was  empirically set to minimize the following cost function:

\begin{equation}
    C(\Theta) = P_{pd} \times C_{miss} \times P_{miss}(\Theta) + P_{th} \times C_{fa} \times P_{fa}(\Theta)
\end{equation}
%$(P_{pd} * C_{miss} * P_{miss}) + (P_{th} * C_{fa} * P_{fa})$.
% Rubén - where P_{pd} and P_{th} denote ..., C_{miss} and C_{fa} are the ..., and P_{miss} and P_{fa} denote..., respectively.
where $P_{pd}$ and $P_{th}$ denote the patient and therapist class prior, $C_{miss}$ and $C_{fa}$ are the cost of miss and false alarm errors, and $P_{miss}$ and $P_{fa}$ denote the system miss and false alarm rate for a specific decision threshold, respectively.
In this work, we are specially concerned with the removal of the therapist's speech, while, up to some extent, we can afford to eliminate some  data of patient's speech. %In fact, our goal is to have clean recordings of the patient group in the OFF state, while we think that 
For these reasons, we chose the following costs and probabilities: $C_{miss}=1$, $C_{fa}=50$, $P_{pd}=0.99$, $P_{th}=0.01$.
\end{comment}

\subsection{Feature extraction}
\label{section:feats}

After the removal of pauses and therapists' interventions, all the clean speech recordings are parameterized in a frame basis. Three different feature sets were used in this work. The first two sets, henceforth referred to as MFCC and MFCC+$\Delta$s, consist of 13 Mel-Frequency Cepstral coefficients (MFCC) in the former case, augmented with the corresponding 13 delta coefficients in the latter. These features were extracted with HTK \cite{HTKBook} using a window size of 25 ms and a window period of 10 ms. The third feature set, referred to as eGeMAPS, consists of the 26 low-level descriptors (LLD) of the eGeMAPS feature set \cite{Eyben:2016}. This is a compact set of acoustic-prosodic features, well-known for their usefulness in a wide range of paralinguistic tasks. These features were extracted with the openSMILE toolbox \cite{Eyben2013} using a frame period of 10 ms and window sizes of 20 and 60 ms depending on the specific LLD \cite{Eyben:2016}. The feature vectors are finally normalized (zero mean and unit variance) on a per-speaker basis. %file

\subsection{Medication state modeling}
The task of assessing PD patients' medication state (ON-OFF) from their speech can be stated as a binary classification problem. 
In a preliminary phase, we experimented a simple model like GMMs, which provided poor results. 
Thus, we use  deep neural networks (DNN) for assigning every input frame in the patient's utterance with a probability of that speech sample corresponding to either states ON or OFF. A final decision for the whole utterance is obtained as the mean value of the individual probabilities. One fundamental aspect in the definition of the approach followed to model the medication state regards the dependency of this task on the speakers' personal characteristics. This comes down to determining whether there exists a set of acoustic-prosodic cues shared by speakers with different gender, ages, and health conditions (including PD development stage) that can be leveraged to develop robust models for this task. Thus, we conducted a preliminary series of experiments in a speaker-independent setup.
\begin{comment}
Thus, we initially conducted experiments in a speaker-independent setup.
\end{comment}
Several machine learning techniques (DNN and SVM) and feature types (MFCC, eGeMAPS, i-vectors and X-vectors) were used to train speaker-independent models that achieved limited performance results up to 65\% utterance-level accuracy. These results, along with the nature of the task itself, justify the adoption of the speaker-dependent approach that is presented in detail below.

\begin{table*}[ht!]
\centering
\caption{Optimal model configurations and results (utterance-level Acc - \%) for speaker-dependent 
medication state assessment.}
\label{tbl:config_res_spkrdep}
\begin{tabular}{lccccccc}
\hline
\textbf{Feature  set} & \textbf{\#Coefficients} & \textbf{Context} & \textbf{Input dim.} & \textbf{Architecture} & \textbf{$\alpha$} & \textbf{Acc devel.} & \textbf{Acc test} \\ \hline
MFCC                & 13 & 15 & 195 & 256, 128, 32, 1 & 0.01 & 93.92 & 88.74 \\
MFCC+$\Delta$s      & 26 & 11 & 286 & 256, 128, 1 & 0.01 & 93.24 & 89.86 \\
eGeMAPS             & 23 & 15 & 345 & 512, 128, 1 & 0.001 & 95.95 & \textbf{90.54} \\ \hline
MFCC+PCA            & 13 & 15 & 95 & 512, 128, 1 & 0.03 & 91.89 & 90.09 \\
MFCC+$\Delta$s+PCA  & 26 & 11 & 85 & 128, 64, 1 & 0.03 & 91.89 & 89.41 \\
eGeMAPS+PCA         & 23 & 15 & 70 & 128, 64, 1 & 0.03 & 96.62 & 87.84 \\ \hline
\end{tabular}
\end{table*}

\subsubsection{Data partitioning}
The pre-processed FraLusoPark dataset is divided into three different subsets for training, development (validation) and test purposes. In a speaker-dependent approach, all these three subsets contain recordings from all the 74 PD patients. Namely, we divide the pre-processed dataset by speech tasks, as following:
\begin{itemize}
\vspace{-0.05cm}\item Train: MPT, DDK, Reading 10 words, Reading prosodic sentences, and Conversation.
\vspace{-0.1cm}\item Development: Reading 10 sentences.
\vspace{-0.1cm}\item Test: /a/, Reading text, and Story telling.
\end{itemize}\vspace{-0.05cm}
The training set contains 740 files (10 files per patient, 5 ON and 5 OFF) and 3983327 samples, which amounts to 55.6$\%$ of the whole pre-processed dataset in terms of utterances. The development set contains 148 files (2 per patient) and 217956 samples (11.1$\%$ of the dataset). The test set contains 444 files (6 per patient) and 1166119 samples, (remaining 33.3$\%$ of the dataset). On average, there are approximately 53800, 2900 and 15800 samples per speaker in these 3 subsets (around 9, 0.5 and 2.6 minutes of speech), respectively. Overall, the number of samples corresponding to the ON and OFF states is well-balanced (2629635 vs. 2737767), although we could observe some notable differences for some speakers, specially in the training subset.

\subsubsection{Medication state classifiers}
Different architectures of feedforward networks were assessed for the goal of learning the relationships between input feature vectors and ON-OFF medication states. Specifically, architectures with 1 to 3 hidden layers with 32 to 512 nodes were validated. ReLU and sigmoid activation functions are used in the hidden and output layers, respectively. The learning rate $\alpha$ is validated, whereas a moderate fixed batch size (32) is used to limit the search space in the validation process. Also, different input context sizes were assessed in our experiments (1, 5, 11, and 15 frames), together with the application of Principal Component Analysis (PCA) for dimensionality reduction. The selection of the optimal values for the free (hyper)parameters was performed based on validation results using the development subset. The final results are computed on the test subset. All the results in this work are reported in terms of accuracy (Acc) at a file/utterance-level.%(model architecture, input context size and learning rate)

In our speaker-dependent approach, the final system consists of 74 individual models, one for each speaker. However, they all share the same configuration (architecture and training hyperparameters). The reason behind this is that just 2 speech recordings from each patient are available for validation purposes. Thus, validation on a per-speaker basis is not feasible. Instead, the average accuracy on the validation subset across all the patients is used to select the optimal global configuration that is then used to train 74 individual DNNs, one for each patient. The DNN classifiers were trained on several GPUs using TensorFlow \cite{tensorflow2015-whitepaper}. %, due to the limited -although very reasonable in this scenario- size of the FraLusoPark dataset

Table \ref{tbl:config_res_spkrdep} shows the optimal configuration for each of the six parameterizations used in this work: the three sets described in Section \ref{section:feats} with and without PCA. The number of coefficients per speech frame are 13, 26 and 23 for MFCC, MFCC+$\Delta$s and eGeMAPS, respectively. A context length of 15 frames was found optimal for the MFCC- and eGeMAPS-based systems, whereas the MFCC+$\Delta$s(+PCA) systems use a context of 11 frames. For the PCA-based systems, the input dimension corresponds to the number of principal components kept, which is determinastically selected as the n largest eigenvalues that explain a 95\% of the total variance in the original input space.%\hl{This leads to a reduction of the input dimension of 51.3\% (MFCC), 70.3\% (MFCC+$\Delta$s) and 79.7\% (eGeMAPS).} %of the covariance matrix of the input data
%\footnote{https://scikit-learn.org/stable/modules/generated/sklearn.decomposition.PCA.html}

\section{Experimental results and analysis}
\label{section:exp}
Table \ref{tbl:config_res_spkrdep} shows the validation and test results (utterance-level accuracies) of the DNN-based systems described above. These results reveal that the medication condition (ON-OFF) of the speakers can be inferred from the acoustic-prosodic information in their speech by means of automatic speech processing and machine learning procedures. The MFCC-based systems achieve accuracies of 88.74\% and 90.09\% in the test set, while the MFCC+$\Delta$s-based systems achieve comparable results of 89.86\% and 89.41\%. One reason to explain the competitive results achieved by the MFCCs alone may be the use of a larger input context size (15 frames), which compensates for the dynamic information provided by the delta MFCC features. The eGeMAPS-based systems achieve accuracies of 90.54\% and 87.84\%.
\begin{comment}
lower accuracies (90.54\% and 87.84\%), despite the complementary prosodic and voice-quality cues provided by this feature set. This issue deserves further analysis since, a priori, that information would seem to be very valuable for the assessment of the medication state in PD patients
\end{comment}
The use of PCA leads to better results for the MFCC-based system, even with a substantial reduction in the dimensionality (51.3\%). For the MFCC+$\Delta$s and the eGeMAPS-based systems the results are 
\begin{comment}
slightly or clearly 
\end{comment}
lower, however. We hypothesize that this may be due to the extreme dimensionality reduction in these two cases (70.3\% and 79.7\%, respectively)
\begin{comment}
and will experiment with a higher number of PCA components in the future
\end{comment}
. 
Finally, it is worth mentioning that the medication state is not equally assessed for all the patients. Since all of them contribute the same number of test utterances, the results in Table \ref{tbl:config_res_spkrdep}  coincide with the average per-speaker accuracies. The speaker variability, expressed in terms of the standard deviation of the per-speaker accuracies, ranges from 12.51 to 14.67.%, which opens the door for a continuous telematic control of the medication effectiveness in PD patients

The markedly different nature of the various speech tasks in the test subset justifies a separate analysis of the results in a per-task basis. Table \ref{tbl:results_spkrdep_task} shows how the performances of the six models depend on the speech task: phonation of the vowel /a/, reading of a short text, and guided storytelling.
\begin{comment}to a great extent, being in general lower on the /a/ task (accuracies around 83.78\% to 89.19\%)
\end{comment}
The best results are achieved on the semi-spontaneous visually-guided storytelling task (accuracies between 92.57 and 95.27\%), followed by the reading of short text task (accuracies between 86.49 and 91.22\%) and the /a/ task (accuracies around 83.78\% to 89.19\%). It is specially remarkable the performance of the eGeMAPS on the story telling task, with a 95.27\% accuracy. 
\begin{comment}
The rightmost column shows the test performance excluding the files corresponding to the /a/ task.
\end{comment}
As a conclusion, the results shown in Table \ref{tbl:results_spkrdep_task} point to the benefit of using natural, (semi)spontaneous speech recordings for the automatic assessment of PD patients medication state.  These observations are consistent with previous work on the relevance of speech tasks for automatic detection of Parkinson's disease \cite{pompili2017automatic}. Overall, the results reported in this work have the potential to contribute to opening new avenues for the automatic distant monitoring of PD patients.
\begin{table}[h!]
\centering
\caption{Test results (utterance-level Acc - \%) for speaker-dependent medication state assessment by task.}
\label{tbl:results_spkrdep_task}
\begin{tabular}{lccc}
\hline
\textbf{Feature  set} & \textbf{/a/} & \textbf{Reading} &  \textbf{Story}  \\
 &  & \textbf{text} & \textbf{telling}  \\ \hline
MFCC                & 86.49 & 87.16 & 92.57  \\
MFCC+$\Delta$s      & 85.14 & 91.22 & 93.24  \\
eGeMAPS             & 89.19 & 87.16 & \textbf{95.27}  \\ \hline
MFCC+PCA            & 86.49 & 89.19 & 94.59  \\
MFCC+$\Delta$s+PCA  & 85.14 & 89.19 & 93.92  \\
eGeMAPS+PCA         & 83.78 & 86.49 & 93.24 \\ \hline
\end{tabular}
\end{table}

\section{Conclusions and future work}
\label{section:conclusions}
This work presents an approach that combines speech processing and deep learning techniques to perform automatic classification of the medication state of PD patients by leveraging personal speech-based bio-markers. We devise a speaker-dependent approach and investigate the relevance of different acoustic-prosodic feature sets. Test results show an accuracy of 90.54\% in a mixed-speech task and a remarkable accuracy of 95.27\% in a semi-spontaneous speech task. These results show the potentials of this approach towards the development of personalized and reliable systems for a daily, remote monitoring and scheduling of medication intake of PD patients.

Further work will pursue the acquisition of more speech data to perform more exhaustive and relevant experimentation, together with an analysis of how the proposed approach scales to new and larger corpora. Presumably, this would also open the door to new and more elaborated approaches using convolutional and/or recurrent neural networks, and to speaker-independent approaches. We plan to continue investigating that approach in combination with the use of speech embeddings (e.g., i-vectors and X-vectors) that may convey specific information relevant to the medication state classification task. Speaker-independent models may also serve as a good starting point for the development of more powerful speaker-dependent models by means of retraining them using speaker-specific speech data.

\bibliographystyle{IEEEtran}

\bibliography{mybib}

\end{document}